\documentclass[superscriptaddress,prd,reprint,showpacs]{revtex4-1}

\usepackage{amsmath,empheq}
\usepackage{amsfonts}
\usepackage{amsthm}
\usepackage{amssymb}
\usepackage{graphicx}
\usepackage{hyperref}
\usepackage{soul}
\usepackage{url}
\usepackage{color}
\usepackage{bm}
\usepackage{cleveref}

\usepackage{chngcntr}
\counterwithout{table}{section}

\graphicspath{ {./images/} }

	\newcommand{\ncd}{\newcommand}
	\ncd{\mrm}    {\mathrm}
	\ncd{\beq} {\begin{equation}}
	\ncd{\eeq} {\end{equation}}

	\def\d{{\rm d}}

	\def\p{{\bf p}}
	\def\q{{\bf q}}

\begin{document}
	\title{Geometric integrator for simulations in the canonical ensemble}
	\date{\today}

	\author{Diego Tapias}
	\email{diego.tapias@nucleares.unam.mx}
	\affiliation{Departamento de F\'isica, Facultad de Ciencias, Universidad Nacional Aut\'onoma de M\'exico, Ciudad Universitaria, Ciudad de M\'exico 04510, Mexico}

	\author{David P. Sanders}
	\email{dpsanders@ciencias.unam.mx}
	\affiliation{Departamento de F\'isica, Facultad de Ciencias, Universidad Nacional Aut\'onoma de M\'exico, Ciudad Universitaria, Ciudad de M\'exico 04510, Mexico}
	\affiliation{Computer Science and Artificial Intelligence Laboratory,  Massachusetts Institute of Technology, 77 Massachusetts Avenue, Cambridge, MA 02139, USA}

	\author{Alessandro Bravetti}
        \email{alessandro.bravetti@iimas.unam.mx}
	\affiliation{Instituto de Investigaciones en Matem\'aticas Aplicadas y en Sistemas, Universidad Nacional Aut\'onoma de M\'exico,
	Ciudad Universitaria, Ciudad de M\'exico 04510, Mexico}

		\begin{abstract}

In this work we introduce a geometric integrator for molecular dynamics simulations of physical systems in the canonical ensemble. In particular,
we consider the equations arising from the so-called density dynamics algorithm with any possible type of thermostat and provide an integrator that preserves the invariant distribution.
Our integrator thus constitutes a unified framework that allows the study and comparison of different thermostats and of their influence on the equilibrium and non-equilibrium (thermo-)dynamic properties of the system. 
To show the validity and the generality of the integrator, we implement it with a second-order, time-reversible method and apply it to the simulation of a Lennard-Jones system with three different thermostats, obtaining good conservation of the geometrical properties and recovering the expected thermodynamic results.
		
	\end{abstract}

%\pacs{04.70.Bw, 04.70.Dy, 02.40.ky}

        \maketitle

 \section{Introduction}
While dynamics in the microcanonical ensemble is easily generated from the standard Hamilton equations of motion for a conservative system,
a problem arises when trying to generate deterministic equations of motion that reproduce a given system in contact with a thermal reservoir at a fixed temperature,
i.e.~a system in the canonical ensemble. This problem is relevant, for example, for molecular dynamics simulations, since the canonical ensemble
is more useful for comparison with experimental situations~\cite{allen1989computer,rapaport2004art,frenkel1996understanding}.
Several algorithms for simulations in the canonical ensemble have been suggested in the literature~\cite{TuckermanBook,leimkuhler2015molecular}; 
here, we focus on those that extend the physical phase space of the system of interest
by adding an extra dimension which takes into account the interaction of the system with the environment. Such methods are usually referred to as thermostat algorithms,
the most well-known being \emph{Nos\'e-Hoover dynamics}~\cite{evans1985thermostat, hoover1985canonical}.

It is known that several possible inequivalent deterministic thermostats correctly generate the canonical ensemble in the physical phase space~\cite{jellinek1988generalization, jellinek1988dynamics, kusnezov1990canonical}. 
Fukuda and Nakamura~\cite{fukuda2002tsallis} provided an algorithm called \emph{Density Dynamics} (DD) that can generate, in principle, any invariant distribution in the extended and in the physical phase space. 
For example, Nos\'e-Hoover dynamics corresponds, in this scheme, to prescribing a Gaussian distribution for an extended variable that controls the interaction between the system and the environment. 
However, this choice is not unique: choosing a different distribution for the extended variable can also lead to a canonical ensemble in the physical phase space.

%Moreover, they also constructed a general scalar invariant of the DD equations of motion that is useful in order to keep track of the correctness of the numerical integration 
%  \cite{fukuda2006construction}.
Recently, two of the present authors introduced a similar algorithm, \emph{Contact Density Dynamics} (CDD)~\cite{bravtap2016}, in which the equations of motion stem from contact Hamiltonian mechanics \cite{bravcruztapias2016}. One can show that this algorithm leads to results equivalent to those of DD; see Section~\ref{DDcanonical}.

Given the fact that a canonical ensemble in the physical phase space can be reproduced by means of different types of thermostatting dynamics, an interesting question
arises as to whether one can distinguish between such thermostats and establish criteria to prefer one over the other, at least for particular systems.
For example, one such criterion is the lack of ergodicity for some value of the thermostatting parameter or for different temperature ranges, which would lead to a dynamics that does not correctly reproduce
the expected thermodynamic properties. Therefore, for instance, one would like to show that for a given system a thermostat is superior over another in that  
it produces the correct results for a wider range of the thermostatting parameter and of the external temperature.
Such a problem can be addressed either with theoretical investigations or with reference to numerical simulations; for the latter, for consistent comparisons it is necessary to have a unified and general framework for the numerical integration of any thermostat dynamics. Moreover, this unique numerical integrator should be constructed to respect the relevant geometric structures of
each different thermostat.

In this work we introduce a numerical integrator with the above prescribed properties. 
It is a geometric integrator (geometric in the sense of \emph{structure-preserving}~\cite{ezra2006integrators})  
for the equations of motion of Density Dynamics, adapted to the simulation of a physical system in the canonical ensemble. 
We show that our integrator works for any choice of the thermostatting mechanism and thus applies to any canonically thermostatted dynamics. 
The integrator is constructed by splitting the vector field of the dynamics and composing the individual flows using the symmetric 
Trotter factorization~\cite{mclachlan2002splitting, mclachlan2006geometric, leimkuhler2015molecular, hairer2006geometric, TuckermanBook}.

To show the usefulness of our integrator, we use it to simulate a Lennard-Jones fluid with $256$ particles in the canonical ensemble within three different thermostatting schemes:
the Gaussian (Nos\'e-Hoover) model, the quartic model of Fukuda and Nakamura~\cite{fukuda2006construction}, and the logistic model that we proposed in~\cite{bravtap2016}.
The results show that our integrator is suitable for performing numerical simulations with any of the thermostatting schemes, and that 
it maintains control over the invariant quantity throughout the simulation once equilibrium is reached. Furthermore, all three schemes generate results that 
are consistent with the canonical ensemble. This is of particular interest to perform simulations where the Nos\'e-Hoover method fails, or takes a long time to reach equilibrium, such as occurs for the harmonic oscillator \cite{hoover1985canonical, cho1992ergodicity, watanabe2007ergodicity, legoll2007nonergodicity},
and to compare the results obtained by different thermostatting schemes. 
%whenever experimental data are not at hand.

We remark that
the main difference between our integrator and the one described by Fukuda and Nakamura is that
in~\cite{fukuda2006construction} the authors built a numerical integrator in the $(2n+2)$-dimensional extended phase space, assuming
that an unspecified splitting was given for the $(2n+1)$-dimensional vector field. 
Here, instead, we directly provide, for the first time, an explicit splitting for the $(2n+2)$-dimensional vector field.

\section{Density Dynamics in the canonical ensemble}
\label{DDcanonical}

The standard Hamilton equations imply the conservation of energy and consequently they generate a dynamics in the microcanonical ensemble. There are several proposals to produce non-Hamiltonian dynamics consistent with different ensembles \cite{hoover1985canonical, kusnezov1990canonical,tuckerman1992chains, leimkuhler2015molecular}. In this work, we consider the Density Dynamics (DD) algorithm, which provides dynamical equations that generate an arbitrary density $\rho(\p,\q,\zeta)$ as the invariant distribution in a $(2n+1)$-dimensional phase space~\cite{fukuda2002tsallis}. 

The DD equations are given by  
\begin{empheq}{align}
	\label{DD1}
	& \dot {q}_{i} =  \frac{\partial\Theta(\p,\q,\zeta)}{\partial p_{i}}\,,\\
	\label{DD2}
	& \dot{p}_{i} =  -\frac{\partial\Theta(\p,\q,\zeta)}{\partial q_{i}}-\frac{\partial\Theta(\p,\q,\zeta)}{\partial \zeta}\, {p}_{i} \,,\\
	\label{DD3}
	& \dot{{\zeta}} =\sum_{i=1}^{n}p_{i} \frac{\partial\Theta(\p,\q,\zeta)}{\partial p_{i}}-{n}\,,
\end{empheq}
where $\p$ and $\q$ are the mechanical variables of the physical system of interest 
and $\zeta$ is an additional variable associated with the degrees of freedom of the thermal reservoir. 
Here $\Theta(\p,\q,\zeta)=-{\rm ln} \rho(\p,\q,\zeta)$ and $i$ goes from $1$ to $n$, the number of degrees of freedom of the physical system.
We take $\rho$ to be the product of two independent distributions,
\beq
\label{rhoextended}
\rho=\rho_{\rm sys}(\p,\q)\rho_{\rm th}(\zeta) \, , 
\eeq
where the desired invariant distribution $\rho_{\rm sys}$ for the physical system is obtained by
integrating out the unphysical degree of freedom $\zeta$ over the distribution $\rho_{\rm th}$ of the thermostat.

A further justification for DD can be given in terms of the geometrical setting provided by Contact Density Dynamics (CDD). In this framework, one derives the equations of motion from contact Hamiltonian mechanics \cite{bravcruztapias2016}; see equations~(11)--(13) in \cite{bravtap2016}. To recover  \eqref{DD1}--\eqref{DD3} from CDD one fixes the contact Hamiltonian to be $h = \exp(\Theta/n)$  and rescales time according to   $dt \mapsto h/n \, dt$.

Another important property of the system \eqref{DD1}--\eqref{DD3} is that it has an invariant function~\cite{fukuda2006construction}. 
To see this, it is convenient to extend the $(2n+1)$-dimensional phase space to one with dimensionality $2n+2$,
denoted in this work as the \emph{Extended Phase Space} (EPS), by addition of the new coordinate $\nu$, whose equation of motion is
\beq
\dot{\nu} = -\frac{{\rm{div}}({X})}{n} \, ,
\eeq
where $X$ is the vector field defining the system~\eqref{DD1}--\eqref{DD3}. A straightforward calculation shows that 
\beq
\label{DD4}
\dot{\nu} = \dfrac{\partial\Theta(\p,\q,\zeta) }{\partial \zeta}
\eeq
and that, given a solution $\phi(t)$ of~\eqref{DD1}--\eqref{DD3}, the quantity
\beq
\label{generalinvariant}
I_\phi(t) = \Theta(\phi(t)) + n \nu(\phi(t)) \, .
\eeq		
is constant with respect to time $t$ (i.e., $\dot{I}_\phi \equiv 0$) and thus is an invariant of the flow.

For performing simulations in the canonical ensemble, the distribution $\rho$ is factored out as in \eqref{rhoextended}:
\beq
\label{extendeddistribution}
\rho(\p,\q,\zeta) = \dfrac{\exp(-\beta H(\p,\q))}{\mathcal{Z}}f(\zeta) \, ,
\eeq
with $H(\p,\q)$ the Hamiltonian of the physical system, $\beta$ a constant (the inverse of the temperature of the reservoir),
$\mathcal{Z}$ the partition function and $f(\zeta)$ a distribution for the variable $\zeta$. 
Substituting the joint distribution \eqref{extendeddistribution} into equations \eqref{DD1}--\eqref{DD3}
and using \eqref{DD4}, we get the following equations of motion in the EPS:
	\begin{empheq}{align}
		& \dot{q}^i =  \beta \frac{\partial H(\p,\q)}{\partial p_i} \notag \, , \\
		& \dot{p}_i=  - \beta \frac{\partial H(\p,\q)}{\partial q^i}  + \frac{f'(\zeta)}{f(\zeta)}p_i \notag	\,, \\
		& \dot{\zeta} =  \beta \sum_{i=1}^n\frac{\partial H(\p, \q)}{\partial p_i} p_i - n \, \notag , \\
	        & \dot{\nu} = -\frac{f'(\zeta)}{f(\zeta)} \notag  \, .
	\end{empheq}
By rescaling the vector field as 
	\beq\label{rescaling}
	{X} \mapsto \frac{X}{\beta} \equiv X_{\rm EPS}\,,
	\eeq 
which is equivalent to the scaling in time $t \mapsto \beta t$, 
we get a more natural form for the equations of motion, namely
	\begin{empheq}{align}
	\label{canon1}
	& \dot{q}^i =  \frac{\partial H(\p,\q)}{\partial p_i}  \, , \\
	\label{canon2}
		& \dot{p}_i=  - \frac{\partial H(\p,\q)}{\partial q^i}  + \frac{f'(\zeta)}{\beta f(\zeta)}p_i 	\,, \\
	\label{canon3}		
		& \dot{\zeta} =  \sum_{i=1}^n\frac{\partial H(\p, \q)}{\partial p_i} p_i - \frac{n}{\beta} \, , \\
		\label{canon4}
	    & \dot{\nu} = -\frac{f'(\zeta)}{\beta f(\zeta)}  \, .
	\end{empheq}

Since $\beta$ is constant, such a rescaling in time changes neither the equilibrium properties 
of the system nor the integral curves of the vector field~\footnote{The situation is more delicate when the rescaling is due to a scalar function (see e.g. \cite{jellinek1988dynamics,fukuda2015double}). However, one can prove that the average results coincide provided one uses the correct rescaled measure~\cite{fukuda2015double}.}.

The invariant quantity~\eqref{generalinvariant} after rescaling reads
	\beq
	\label{invariantcanonical}
	I_\phi(t) = H(\p(\phi(t)),\q(\phi(t)) - \dfrac{{\rm{ln}} f(\zeta(\phi(t))}{\beta} + \frac{n}{\beta} \nu(\phi(t)) \, .
	\eeq	

%{\bf Without taking into account any conservation law}, 
Finally, it can be shown that the dynamics~\eqref{canon1}--\eqref{canon4} in the EPS has the invariant measure~\cite{Tuckerman2001}
        \beq
        \label{generalmeasure}
        \d\mu  = e^{n \nu} \d^n q \d^n p \d \zeta \d \nu \, ,
        \eeq
        which, using the conservation law~\eqref{invariantcanonical}, can be rewritten as
        \beq
        \label{measurewithconservation}
        \d\mu = e^{n \nu} \delta \left( H(\p, \q) -  \dfrac{{\rm{ln}} f(\zeta)}{\beta} + \frac{n}{\beta} \nu - C \right)  \d^n q \d^n p \d \zeta \d \nu \,,
        \eeq
where $C$ is a constant defined by the initial condition.
From~\eqref{measurewithconservation} a canonical distribution for the physical phase space follows after integration of the additional degrees of freedom $\zeta$ and $\nu$.
Therefore, assuming ergodicity and provided there are no additional integrals of motion, 
the dynamics~\eqref{canon1}--\eqref{canon4} dynamically generates the canonical ensemble~\cite{Tuckerman2001}.

We remark that the role of the thermostat is codified in the function $f(\zeta)$ and that different choices of $f$ lead to different dynamics, which are all,
in principle, consistent with the canonical ensemble. For example, choosing $f(\zeta)$ as a Gaussian distribution one obtains the Nos\'e-Hoover dynamics,
choosing a quartic distribution one obtains the dynamics proposed in~\cite{fukuda2006construction},
and with a choice of a logistic distribution one obtains the dynamics introduced in~\cite{bravtap2016}.

\section{Geometric Integration}
In this section, we present the main result of this paper, i.e.~a geometric numerical integrator for the flow~\eqref{canon1}--\eqref{canon4} that applies
 to any possible choice for the thermostatting function $f(\zeta)$.

\subsection{Splitting of the vector field}
An integrator is called \emph{geometric} if it preserves exactly 
one or more of the geometric properties of a system of differential equations~\cite{mclachlan2006geometric}. 
For example, symplectic integrators are suitable for the integration of systems in the microcanonical ensemble, 
as they conserve the symplectic form and hence the phase space volume exactly~\cite{mclachlan2006geometric}.
Furthermore, even though the energy is not preserved exactly by the approximate dynamics, the errors 
are bounded along the integration; see~\cite{leimkuhler2015molecular} for details. 
So, it is a desirable property of an integrator to be geometric or structure-preserving \cite{ezra2006integrators}.

We now present a geometric integrator for the system~\eqref{canon1}--\eqref{canon4} with properties analogous to those of symplectic integrators 
for conservative systems, since it conserves the invariant measure \eqref{generalmeasure} and maintains 
the error in the invariant quantity \eqref{invariantcanonical} bounded. 
The fulfillment of the first property can be examined analytically, while the second property is checked through numerical experiments.

To build the integrator we use the \emph{splitting method}~\cite{mclachlan2002splitting}. 
The idea of this method is to decompose a vector field $X$ into vector fields $X_i$, with $X = \sum_i X_i$,
such that each $X_i$ 
shares the same properties as $X$; 
the advantage is that the $X_{i}$s are easier to treat numerically. 
In our case, the property that we want to preserve with the splitting is the conservation of the invariant measure.
For concreteness, since $X_{\rm EPS}$ given in~\eqref{rescaling} satisfies
	\beq
	\pounds_{X_{\rm EPS}} \d\mu = 0 \, ,
	\eeq
where $\pounds$ is the Lie derivative (i.e.~the derivative along the vector field $X_{\rm EPS}$~\cite{leimkuhler2015molecular}) 
and $\d \mu$ is given in~\eqref{generalmeasure}, 
we require each vector field in the splitting $X_{\rm EPS}~=~\sum_i X_i$ to separately satisfy 
	\beq
	\label{invariance}
	\pounds_{X_i} \d\mu = 0 \, .
	\eeq

We begin by separating $X_{\rm EPS}$ into two parts,
	\beq
	\label{generalsplitting}
	X_{\rm EPS} = X_{\rm sys} + X_{\rm th}\,, 
	\eeq
with $X_{\rm sys}$ the vector field  associated with the Hamiltonian evolution of the physical system,
	\begin{align}
	X_{\rm sys} &= X_{\rm sys}^{1} + X_{\rm sys}^{2}\label{splittingsys} \\
	&=  \left[- \frac{\partial H(\bf{p},\bf{q})}{\partial q^i}  \frac{\partial}{\partial {p_i}} \right] + \left[ \frac{\partial H(\bf{p},\bf{q})}{\partial p_i} \frac{\partial}{\partial {q^i}} \right] 
	\end{align}
and $X_{\rm th}$ the vector field associated with the action of the thermostat,
	\begin{align}
	\begin{split}
	X_{\rm th} {}& = X_{\rm th}^{1} + X_{\rm th}^{2} \label{splittingth}
	\end{split} \\
	\begin{split}
	{}& =  \left[ \left(\sum_{i=1}^n \frac{\partial H(\bf{p},\bf{q})}{\partial p_i} p_i - \frac{n}{\beta} \right) \frac{\partial}{\partial {\zeta}} \right] \\ 
	{}& + \left[\frac{g(\zeta)}{\beta}p_i \frac{\partial}{\partial {p_i}}  - \frac{g(\zeta)}{\beta} \frac{\partial}{\partial \nu} \right]\,,
	\end{split}
	\end{align}
where $g(\zeta) \equiv {\d {\rm ln} f(\zeta)}/{\d \zeta}$. 
It is easy to check that if the Hamiltonian of the physical system is of the form $H(\p, \q) = K(\p) + V(\q)$, then
	\beq
	\pounds_{X_{\alpha}^{(i)}} \d \mu = 0 \quad \text{for} \quad \alpha = {\rm sys, th} \quad \text{and} \quad i = 1,2 \, 
	\eeq
and thus the individual vectors $X_{\rm sys}^{1}, X_{\rm sys}^{2}, X_{\rm th}^{1}$ and $X_{\rm th}^{2}$ provide the required splitting of $X_{\rm EPS}$.

We proceed with the composition of the individual flows generated by each vector field $X_{\alpha}^i$. 
There are different ways to perform such a composition; 
see e.g.~\cite{mclachlan2002splitting, mclachlan1995numerical}. 
Here, we use a simple factorization scheme that is enough to illustrate the validity of our integrator.

\subsection{Numerical algorithm}
We recall that formally the flow $\phi_{\rm EPS}(t)$ defined by the vector field $X_{\rm EPS}$ is applied to an initial condition $\omega$ as~\cite{leimkuhler2015molecular}
	\beq\label{applyflow}
	\phi_{\rm EPS}(t)(\omega) = \exp(tX_{\rm EPS})(\omega) \, .
	\eeq
Since the right hand side of~\eqref{applyflow} cannot be evaluated analytically, 
we take advantage of the splitting of the previous section and of a particular method of composition in order to numerically integrate the dynamics~\eqref{canon1}--\eqref{canon4}. 
The method of composition that we use in this work is the \emph{symmetric Trotter factorization} 
(also known as the Strang splitting formula or Suzuki's 2nd order method)~\cite{TuckermanBook, hairer2006geometric, ishida1998water}, 
which is one of the simplest methods to implement a geometric algorithm and test its properties~\cite{ezra2006integrators}. 

The Trotter factorization of the flow corresponding to the splitting \eqref{generalsplitting}, applied to a single time step $\tau$, is
	\begin{align}
	\begin{split}
	\exp(\tau X_{\rm EPS}) 
	&= \exp\left(\frac{\tau}{2}X_{\rm th} \right) \exp(\tau X_{\rm sys} ) \exp\left(\frac{\tau}{2}X_{\rm th} \right)\\ 
	&+ \mathcal{O}(\tau^3)\,.
	\end{split}  \raisetag{1.\baselineskip}
	\label{ec25}
 	\end{align} 
Applying again the Trotter factorization to the further splittings~\eqref{splittingsys} and~\eqref{splittingth}, we obtain
	\beq
	\begin{split}
	\exp(\tau X_{\rm EPS}) = &\exp\left(\frac{\tau}{4} X_{\rm th}^{1} \right) \exp\left(\frac{\tau}{2} X_{\rm th}^{2} \right) \exp\left(\frac{\tau}{4} X_{\rm th}^{1} \right) \\ &\exp\left(\frac{\tau}{2} X_{\rm sys}^{1} \right) \exp(\tau 
	X_{\rm sys}^{2}) \exp \left(\frac{\tau}{2} X_{\rm sys}^{1} \right) \\ &\exp\left(\frac{\tau}{4} X_{\rm th}^{1} \right) \exp\left(\frac{\tau}{2} X_{\rm th}^{2} \right) \exp\left(\frac{\tau}{4} X_{\rm th}^{1} \right) \\ &+ \mathcal{O}
	(\tau^3) \, .
	\end{split} \raisetag{1.\baselineskip}
	\label{flow}
	\eeq
Alternatively, the superscripts $1$ and $2$ may be exchanged, producing an equivalent algorithm. 

From now on, we consider a mechanical system with Hamiltonian 
	\beq
	 H(\mathbf{p},\mathbf{q})= \sum_{i=1}^{n} \frac{p_i^2}{2m_i} + V(\bf{q})\, .
	\eeq
The flow defined by \eqref{flow} applied to any initial condition $\omega$ starts with the evaluation of the operator $\exp\left(\frac{\tau}{4}X_{\rm th}^{1} \right)$, namely the translation
	\begin{widetext}
	\begin{align}
	\label{thermo1}
	\exp\left(\frac{\tau}{4}X_{\rm th}^{1} \right) (\omega) = \exp\left(\frac{\tau}{4} \left(\sum_{i=1}^n\frac{p_i^2}{m_i}  - \frac{n}{\beta} \right) \frac{\partial}{\partial {\zeta}} \right) \begin{pmatrix}
	q^i \\
	p_i \\
	\zeta \\
	\nu
	  \end{pmatrix} 
	=  \begin{pmatrix}
	 q^i \\
	p_i \\
	\displaystyle \zeta+ \frac{\tau}{4}\left(\sum_{i=1}^n \frac{p_i^2}{m_i}  - \frac{n}{\beta} \right)\\
	\nu
	 \end{pmatrix}\,.
	\end{align}
	\end{widetext}

We proceed to evaluate ${\rm exp}(\frac{\tau}{2}X_{\rm th}^{2})$.
Note that the components of $X_{\rm th}^{2}$ commute with each other.
This allows us to apply directly the operator $\exp(\frac{\tau}{2} X_{\rm th}^{2})$ over the evolved initial condition without a further splitting, to obtain
	\begin{widetext}
	\begin{align}
	\label{thermo2}
	&\exp \left(\frac{\tau}{2}X_{\rm th}^{2} \right) \begin{pmatrix}
	 q^i \\
	p_i \\
	\displaystyle \zeta + \frac{\tau}{4}\left(\sum_{i=1}^n \frac{p_i^2}{m_i}  - \frac{n}{\beta} \right)\\
	\nu
	  \end{pmatrix} = \notag 
	  \exp\left(\frac{\tau}{2} \left( \frac{g(\zeta)}{\beta }p_i \frac{\partial}{\partial {p_i}}  - \frac{g(\zeta)}{\beta} \frac{\partial}{\partial \nu} \right) \right) \begin{pmatrix}
	 q^i \\
	p_i \\
	\displaystyle \zeta + \frac{\tau}{4}\left(\sum_{i=1}^n \frac{p_i^2}{m_i}  - \frac{n}{\beta} \right)\\
	\nu
	  \end{pmatrix} = \notag \\
	&\exp\left(\frac{\tau}{2}\left(\frac{g(\zeta)}{\beta}p_i \frac{\partial}{\partial {p_i}} \right) \right) \begin{pmatrix}
	 q^i \\
	p_i \\
	\displaystyle \zeta + \frac{\tau}{4}\left( \sum_{i=1}^n \frac{p_i^2}{m_i}  - \frac{n}{\beta} \right) \\
	\displaystyle \nu - \frac{\tau}{2}\left(\frac{g(\zeta)}{\beta}\right)
	  \end{pmatrix} = 
	  \begin{pmatrix}
	 q^i \\
	\displaystyle p_i\exp\left(\frac{\tau}{2}\left(\frac{g(\zeta)}{\beta}\right)\right)\\
	\displaystyle \zeta + \frac{\tau}{4}\left(\exp\left(\tau \frac{g(\zeta)}{\beta} \right)\sum_{i=1}^n \frac{p_i^2}{m_i}  - \frac{n}{\beta} \right) \\
	\displaystyle \nu - \frac{\tau}{2}\left(\frac{g(\zeta)}{\beta}\right)
	  \end{pmatrix} \,,
	\end{align}
	\end{widetext}
where in the last equality we have used the identity $[\exp\left(cx\frac{\partial}{\partial x}\right)]f(x) = f(x\exp(c))$~\cite{TuckermanBook}. 
Then we apply again the operator  $\exp\left(\frac{\tau}{4}X_{\rm th}^{1} \right)$ to the evolved condition~\eqref{thermo2}, obtaining 
	\begin{align}
	\label{thermo3}
	  \begin{pmatrix}
	 q^i \\
	\displaystyle p_i\exp\left[\frac{\tau}{2}\left(\frac{g(\tilde{\zeta})}{\beta}\right)\right]\\
	\displaystyle \tilde{\zeta} + \frac{\tau}{4}\left[\exp\left(\tau \frac{g(\tilde{\zeta})}{\beta} \right)\sum_{i=1}^n \frac{p_i^2}{m_i}  - \frac{n}{\beta} \right] \\
	\displaystyle \nu - \frac{\tau}{2}\left(\frac{g(\tilde{\zeta})}{\beta}\right) 
	  \end{pmatrix} \, ,
	\end{align}
with 
	\beq
	\displaystyle \tilde{\zeta} = \zeta + \frac{\tau}{4}\left(\sum_{i=1}^n  \frac{p_i^2}{m_i}  - \frac{n}{\beta} \right)\,.
	\eeq 
The set of equations \eqref{thermo1}, \eqref{thermo2}, \eqref{thermo3} defines the first integral operator $L_{\rm th}(\tau)=\exp(\tau/2X_{\rm th})$ in~\eqref{ec25}, representing the action of the thermostat.

Now we proceed with the composition of the evolution operators of the vector fields associated to the physical system. 
A straightforward calculation shows that in our scheme the factorization
	\beq
	 \exp(\tau X_{\rm sys} ) = \exp\left(\frac{\tau}{2} X_{\rm sys}^{1} \right) \exp(\tau X_{\rm sys}^{2}) \exp \left(\frac{\tau}{2} X_{\rm sys}^{1} \right)
	\eeq
 leads to the \emph{velocity Verlet algorithm}, which is a standard second-order integrator for Hamiltonian systems~\cite{TuckermanBook,leimkuhler2015molecular};
 we write $L_{\rm Verlet}(\tau)= \exp(\tau X_{\rm sys} )$.
 Therefore we conclude that the total effect of the operator on the right hand side of \eqref{ec25} may be codified in the following composition for each time step $\tau$
 and initial condition $\omega$
	\beq\label{mainresult}
	\left[ L_{\rm th}(\tau) \circ L_{\rm Verlet}(\tau) \circ L_{\rm th}(\tau) \right] (\omega)
	\eeq	
which is our geometric algorithm for the integration of the equations~\eqref{canon1}--\eqref{canon4}. We remark that we recover the known integrator for Nos\'e-Hoover dynamics considered in \cite{ishida1998water, ezra2006integrators}.
An extension of this factorization to a higher-order method may be obtained using the Suzuki-Yoshida scheme \cite{TuckermanBook}. 

\section{Numerical experiments}

In this section, we report the results of an implementation of our algorithm in the Julia language (available at \cite{julialennard}), 
to perform simulations of a $256$-particle Lennard-Jones fluid in 3 dimensions. 

\subsection{Thermostat distributions}
We use three different thermostat distributions $f(\zeta)$ %(and consequently three different functions $g(\zeta)$) already proposed in the literature
for the dynamics~\eqref{canon1}--\eqref{canon4}, corresponding to the known different dynamics mentioned in the introduction.
Namely, the Gaussian distribution, which yields the classical Nos\'e-Hoover dynamics, 
the logistic distribution, introduced in the context of Contact Density Dynamics~\cite{bravtap2016} and the quartic distribution, considered by Fukuda and 
Nakamura in Density Dynamics \cite{fukuda2002tsallis}.
They are summarized in Table~\ref{table:3distributions}, together with the single free parameter associated to each one.
	\begin{table}[h!] %%%check space
	\begin{tabular}{|c|c|c|}
	\hline 
	Distribution & $f(\zeta)$ & Parameter \\
	\hline
	Gaussian & $\sqrt{\dfrac{\beta}{2\pi Q}}{\exp\left(-\dfrac{\beta \zeta^2}{2 Q}\right)}$ & ${Q}$ \\ 
	\hline
	Logistic & $\dfrac{\exp(\zeta - m)}{ (1 + \exp(\zeta - m))^2}$ &  $m$ \\ 
	\hline  
	Quartic & $\dfrac{2\Gamma(3/4) }{\sqrt{2}\pi c^{-1/4}}\exp(-c\, \zeta^4)$ & $c$ \\ 
	\hline 
	\end{tabular} 
	\caption{Distributions $f(\zeta)$ used to perform molecular simulations. 
	The parameter in the Gaussian distribution corresponds to the standard deviation, while in the logistic distribution it corresponds to the mean.
	The quartic distribution is a generalized Gaussian distribution with parameter $c$ related to the standard deviation~\cite{nadarajah2005generalized}.}
		\label{table:3distributions}
	\end{table}

\subsection{Methods and notation}

We use a shifted-force potential \cite{allen1989computer} 
with a cutoff distance $r_c^* = 2.5$ (the superscript $^*$ denotes reduced units~\cite{frenkel1996understanding}). 
In addition, we consider the system enclosed in a cubic box with periodic boundary conditions~\footnote{The use of periodic conditions introduces three additional conserved quantities, 
the components of the total linear momentum. By considering an initial condition with vanishing linear momentum, 
we may continue using the DD equations with the corrected number of degrees of freedom $\tilde{n} = n-3$. 
For a discussion concerning the Nos\'e-Hoover case see \cite{melchionna2000constrained, chojoann1993}}. 

For comparison with experimental values, we take this potential as a model for argon.  A time step $\Delta t^* =0.005$ 
(corresponding to a physical one of $\Delta t = 11~\rm{fs}$) and a reduced density $\rho^* = 0.8$ ($\rho = 34~\rm{mol/L}$) are used. $N=256$ is the number of particles considered. 
$T_{\rm sys}^{*}$ refers to the instantaneous reduced temperature of the system
	\beq\label{reducedTsys}
	T_{\rm sys}^{*}=\frac{2 \langle K^{*}\rangle}{3(N-1)}\,,
	\eeq 
while $T_{\rm th}^{*}$ refers to the fixed temperature of the thermostat. 
The conversion factor between the reduced and physical temperature is $119.8~{\rm K}$. 
The simulations are carried out by setting the free parameters in the distributions to the values $Q =1.0$, $m = 2.0$ and $c=0.1$.

Finally, we take an initial configuration with particles arranged in a cubic simulation cell  containing 64 face-centered cubic unit cells
and velocities determined by the Boltzmann distribution at temperature $T^*_{\rm th}$.

\subsection{Results}
\begin{figure}[h!]
	\includegraphics[width=0.5\textwidth]{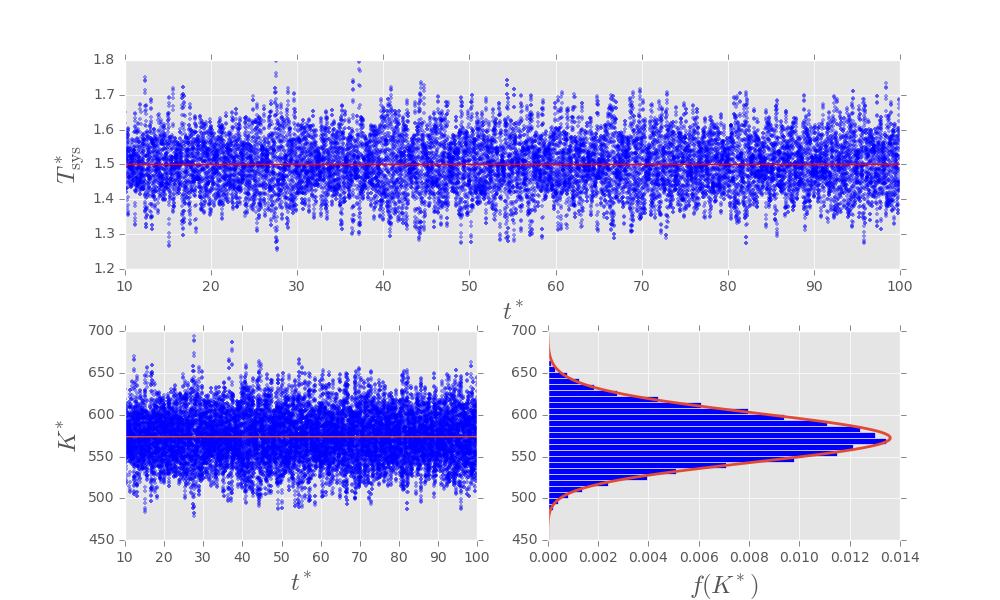} 
        \caption{Evolution of the reduced temperature of the system~\eqref{reducedTsys} at equilibrium with the Gaussian thermostat at a reduced temperature $T_{\rm th}^{*}=1.5$. 
        We report also the behavior of the reduced kinetic energy $K^{*}$ and its numerical distribution $f(K^{*})$, compared with the theoretical one~\eqref{Kdistribution}.}
        \label{Gaussian}
        \end{figure}

\begin{figure}[h!]
        \includegraphics[width=0.5\textwidth]{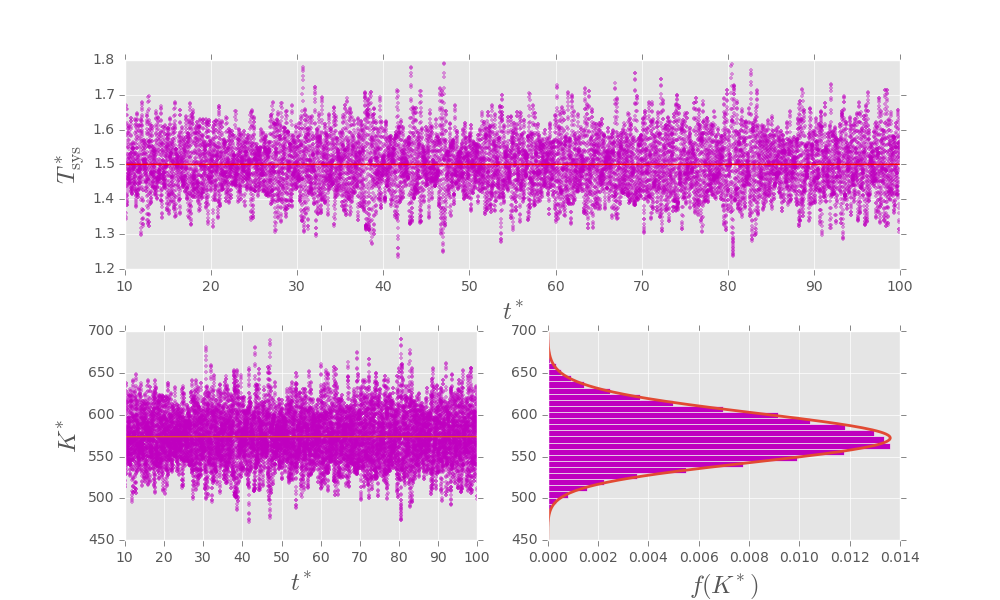} 
       \caption{Evolution of the reduced temperature of the system~\eqref{reducedTsys} at equilibrium with the logistic thermostat at a reduced temperature $T_{\rm th}^{*}=1.5$. 
        We report also the behavior of the reduced kinetic energy $K^{*}$ and its numerical distribution $f(K^{*})$, compared with the theoretical one~\eqref{Kdistribution}.}
               \label{logistic}
        \end{figure}

\begin{figure}[h!]
        \includegraphics[width=0.5\textwidth]{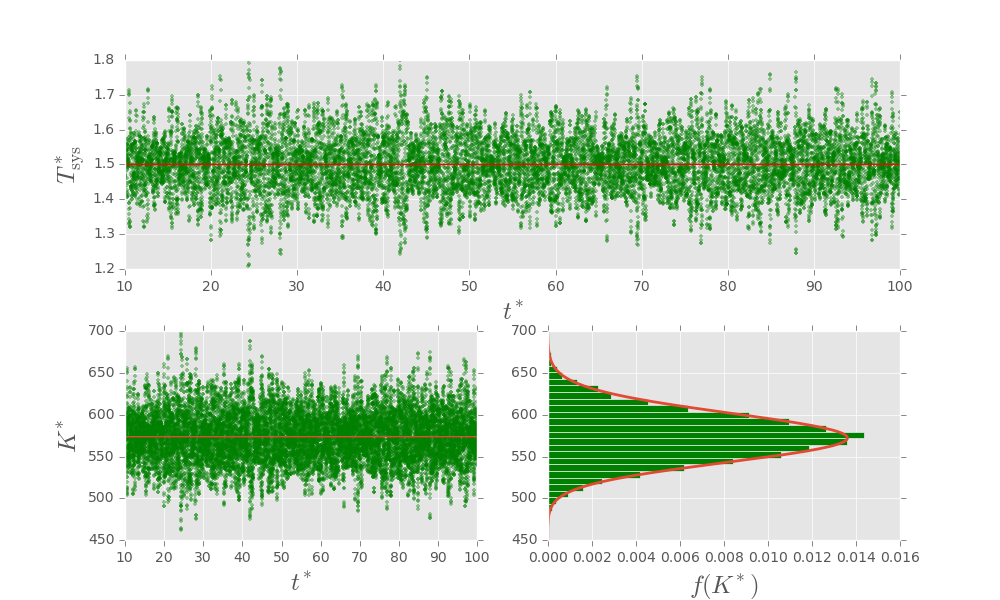} 
        \caption{Evolution of the reduced temperature of the system~\eqref{reducedTsys} at equilibrium with the quartic thermostat at a reduced temperature $T_{\rm th}^{*}=1.5$. 
        We report also the behavior of the reduced kinetic energy $K^{*}$ and its numerical distribution $f(K^{*})$, compared with the theoretical one~\eqref{Kdistribution}.}
                \label{quartic}
        \end{figure}

In figures \ref{Gaussian}, \ref{logistic} and \ref{quartic}, we show the behaviour of $T_{\rm sys}^{*}$ with respect to time 
after equilibrium is reached for the three different thermostats. 
%Each thermostat is set 
%at three different reduced temperatures $T_{\rm th}^{*}$.
It is evident that $T_{\rm sys}^{*}$ fluctuates around a mean value that coincides with $T_{\rm th}^{*}$.
We report also the behavior of the reduced kinetic energy and its numerical distribution compared to the theoretical one,
	\beq\label{Kdistribution}
	\rho(K^{*})=\dfrac{{\rm e}^{-\beta K^{*}}\left(K^{*}\right)^{\frac{3(N-1)}{2}-1}}{\beta^{\frac{3(N-1)}{2}}\Gamma\left(\frac{3(N-1)}{2}\right)}\,,
	\eeq
showing that the results agree with the theoretical expectations in all three cases.

\begin{figure}[h!]
        \includegraphics[width=0.5\textwidth]{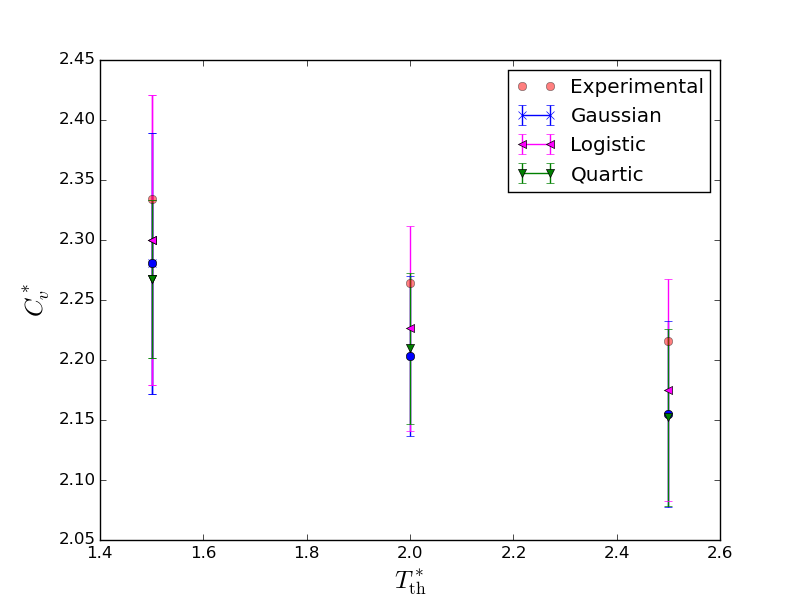} 
        \caption{Reduced heat capacities of each thermostatted dynamics at three different thermostat temperatures ($T_{\rm th}^{*}=1.5$, $T_{\rm th}^{*}=2.0$ and $T_{\rm th}^{*}=2.5$). 
        The error bars represent the standard deviation of the result corresponding to 50 realizations of the simulation.} 
        \label{heatcapacity}
        \end{figure}

The fluctuations of the energy are used to determine the reduced heat capacity of the system according to
	\beq
	C_{v}^{*}=\frac{\langle(H^{*}-\langle H^{*}\rangle)^{2}\rangle}{N\langle T_{\rm sys}^{*}\rangle^{2}}\,,
	\eeq
where the average is performed with respect to time in the numerical simulation.
In figure~\ref{heatcapacity} we
compare the results obtained with the tabulated values for argon under the same conditions~\footnote{NIST standard reference database. Available from: http://webbook.nist.gov/chemistry/fluid/}. 
We display the mean values of $C_{v}^{*}$ for the three different temperatures analyzed, together with the corresponding standard deviation bars obtained using a sample of 50 simulations
for each case.
Clearly our results agree statistically with the experimental values. 

%Though not conclusive, this together with the obtained mean value  are indications that we have correctly generated the canonical ensemble. 

As a final test that the three different dynamics correctly reproduce the canonical ensemble, 
we calculate in each case the scaled covariance of the kinetic and the potential energy, defined as 
\beq
{\rm cov}^*(K, U) = \dfrac{\langle \Delta K^* \Delta U^* \rangle}{N \langle T^*_{\rm sys} \rangle ^2} \,
\label{formulacov}
\eeq
with $\Delta Y = Y - \langle{Y}\rangle$ for  $Y = U^{*}, K^{*}$. This quantity is expected to vanish at equilibrium \cite{valenzuela2015}. 
In Table~\ref{covariances} we display the statistical results of covariances for samples of 50 simulations. They are in agreement with the expectation in all cases.

\begin{table}[h!]
\resizebox{\columnwidth}{!}{
\begin{tabular}{|c|c|c|c| c| c| c|}
\hline 
 {} & \multicolumn{2}{c|}{$T_{\rm th}^* = 1.5$} &  \multicolumn{2}{c|}{$T_{\rm th}^* = 2.0$} &  \multicolumn{2}{c|}{$T_{\rm th}^* = 2.5$} \\ 
\hline 
 Reservoir  & mean & std & mean & std & mean & std  \\
\hline
Gaussian & $-6.22 \times 10^{-4}$ & $0.017$   & $-1.67 \times 10^{-3}$ & $0.014$  & $-6.76 \times 10^{-4}$ &  $0.012$ \\ 
\hline 
logistic & $8.64 \times 10^{-3}$ & $0.025$    &  $5.73 \times 10^{-3}$ & $0.018$ & $2.66 \times 10^{-3}$ & $0.014$   \\ 
\hline 
quartic & $-1.23 \times 10^{-3}$ & $0.015$ & $1.00 \times 10^{-4}$ & $0.013$   & $7.61 \times 10^{-4}$ & $0.008$    \\ 
\hline 
\end{tabular} 
}
\caption{Mean reduced covariances (mean) and standard deviations (std) of the kinetic and potential 
energy calculated with formula \eqref{formulacov} for 50 simulations in each case under the specified conditions.}
\label{covariances}
\end{table}

\begin{figure}[h!]
\includegraphics[width=0.5\textwidth]{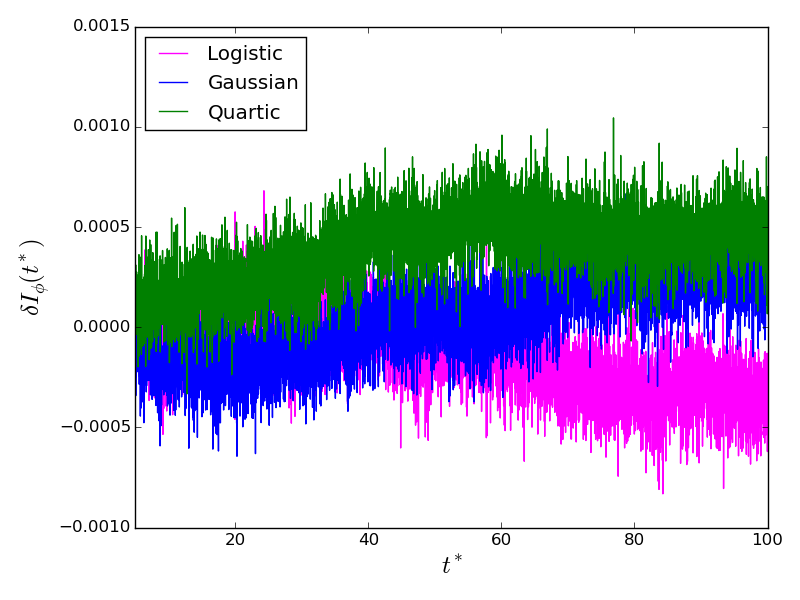} 
\caption{Evolution of the relative error in the invariant $\delta I_{\phi}(t^*)~=~\frac{I(t^*) - I(t_{\rm eq}^*)}{I(t_{\rm eq}^*)}$  after equilibrium has been reached ($t_{\rm eq}^* = 5.0$). 
A random simulation was chosen for the same reduced temperature $T^*_{\rm th} = {1.5}$.} 
%\textbf{Decir que es una simulaci\'on representativa y que no tiene nada de particular respecto a cualquiera que pudieras graficar, m\'as all\'a de empezar a ser calculada en el equilibrio}}
\label{invariantgraph}
\end{figure}

Additionally, to test the validity of the integrator,
we keep track of the evolution of the invariant quantity~\eqref{invariantcanonical} as the system is numerically integrated. 
Since the algorithm is constructed so as to preserve just the invariant measure, 
{the invariant quantity is not necessarily preserved exactly by the approximate dynamics.} 
However,
its fluctuations are expected to be bounded and not to have any particular drift. 
Figure~\eqref{invariantgraph} confirms this expectation for the three thermostat dynamics considered. 
%The fluctuations are small and they do not show any drift. 

\section{Conclusions}
In this work we have provided a new integrator for molecular dynamics simulations in the canonical ensemble.
The main characteristics of our integrator are that it is geometric, in the sense that it preserves the invariant measure of the dynamics,
and that it works for any type of thermostat within the density dynamics formalism. 
Different thermostatted dynamics may be generated according to equations~\eqref{canon1}--\eqref{canon4} by specifying the distribution $f(\zeta)$ and they are all expected to give rise to a canonical ensemble, provided the ergodic assumption is fulfilled. %and there are no other integral of motion
%beside~\eqref{invariantcanonical}. 
Our algorithm provides a unified framework for the integration of all these thermostatted dynamics, and thus a useful tool for their comparison.
In fact, while in principle the thermodynamic properties of the system do not depend on the subtleties of the thermostat algorithm, 
in practice there are issues that may cause a thermostat to be preferable over another. 
As an example, it is known for the Nos\'e-Hoover thermostat that the choice of $Q$ is an important factor for the correctness 
of the simulation~\cite{cho1992ergodicity, valenzuela2015}.
%We explored a range of values of $Q$ in each model, 
%and in this work chose a value of it according to some practical considerations such as a rapid time of equilibration. 
%It deserves further study the allowed range of values of $Q$ for each distribution and even more, the factors which it depends on. 

In order to obtain results that are consistent and comparable for simulations of different thermostatted dynamics, it is necessary to use the same integrator for all cases. 
Our integrator provides such a unified scheme to compare 
different canonical dynamics, as shown here, or even to extend the comparison to more sophisticated deterministic algorithms, such as Nos\'e-Hoover chains~\cite{tuckerman1992chains}.

Future work  will be devoted to the implementation of the algorithm with higher-order methods, to the simulation of several physical systems and to the extension of the integrator to ensembles different from the canonical one.
%A relevant application would be the case in which experimental data are difficult to obtain and therefore a comparison of the results of different thermostatting schemes
%within the same formalism can be used as a check for the results.

\section*{Acknowledgements}
AB is supported by a DGAPA-UNAM postdoctoral fellowship. DT acknowledges financial support from CONACYT, CVU No. 442828. DPS acknowledges financial support from DGAPA-UNAM grant PAPIIT-IN117214, and from a CONACYT sabbatical fellowship.
DPS thanks Alan Edelman and the Julia group at MIT for hospitality while this work was completed.

\bibliographystyle{ieeetr}
\bibliography{GTD}

\end{document}